Presence of Two Maxima in the Isothermal Free-Radical Polymerization Rate of Isobornyl Methacrylate Retarded by Oxygen

**Ezequiel R. Soulé, Julio Borrajo, and Roberto J. J. Williams***

*Institute of Materials Science and Technology (INTEMA), University of Mar del Plata and National Research Council (CONICET), J. B. Justo 4302, 7600 Mar del Plata, Argentina*

* To whom correspondence should be addressed, e-mail: williams@fi.mdp.edu.ar

ABSTRACT: The kinetics of the free-radical polymerization of isobornyl methacrylate (IBoMA) at 80 ºC, initiated by benzoyl peroxide (BPO), was determined by differential scanning calorimetry (DSC), using sample pans previously sealed in air or in nitrogen. The polymerization was arrested by vitrification at a final conversion close to 0.80. The kinetics in nitrogen could be fitted using initiation and propagation rate constants reported in the literature, a functionality of the termination rate constant controlled by translational diffusion based on the free-volume model, and an efficiency factor for the initiator decomposition that decreased with conversion in the medium-high conversion range. The presence of oxygen led to a significant decrease in the polymerization rate, an effect that was enhanced when decreasing the sample size (increasing the surface-to-volume ratio). The original finding was the presence of two maxima in the isothermal polymerization rate, a fact that was confirmed by replicating experiences in two different DSC devices and changing the amount of initiator in the formulation. The first maximum in the polymerization rate was explained by the decrease of the diffusion rate of macromolecular radicals to the oxygen-rich boundary, an effect that started at low conversions. The second maximum was related to the decrease in the solubility/diffusion of oxygen in the reaction medium, a phenomenon that was particularly severe at high conversions.

**Introduction.**

Molecular oxygen exerts a detrimental effect on radical-induced polymerizations because of its high reactivity toward radical species. This effect is particularly pronounced in the UV cure of coatings because of large surface-to-volume ratios, which create convenient conditions for oxygen diffusion into the coating.[1] Radicals are scavenged by oxygen with a rate $r_O = k_O$ [R·][$O_2$], where $k_O \sim 5 \times 10^8$ L mol$^{-1}$ s$^{-1}$.[2] This reaction competes with propagation that takes place at a rate $r_p = k_p$ [R·][M], where $k_p \sim 10^3$ L mol$^{-1}$ s$^{-1}$. The equilibrium concentration of oxygen dissolved in a typical (metha)acrylate monomer lies in the range of [$O_2$] $\sim 10^{-3}$ mol L$^{-1}$,[2] whereas [M] $\sim 5$ mol L$^{-1}$, leading to an initial $r_O/r_p \sim 10^2$. Therefore, dissolved oxygen must be consumed to a high extent before polymerization may be effectively initiated. This occurs during the induction period after which oxygen leads to a retardation effect due to its continuous diffusion from the atmosphere through the surface of the film.

The objective of the present work was to compare the polymerization rate of films of a methacrylic monomer exposed to either a nitrogen atmosphere or air, in the whole range of conversions. As suggested in a recent review,[3] the polymerization was conveniently followed by differential scanning calorimetry (DSC) in the isothermal mode, initiating the reaction by the thermal decomposition of a typical initiator. The use of DSC has two main advantages over other methods: a) it enables to perform the polymerization under strictly isothermal conditions due to the small size of the samples, b) the exit signal may be directly related to the polymerization rate. For the purpose of this study, DSC cells containing the monomer and an initiator were closed either in a nitrogen atmosphere or in air. Kinetic results obtained by DSC were validated by comparison with conversion vs. time curves obtained by Fourier-transformed infrared spectroscopy (FTIR). Isobornyl methacrylate (IBoMA) was selected for this study due to its low vapor pressure (3 torr at 80 ºC),[4] that facilitated the preparation of samples avoiding experimental problems related to mass losses.

The manuscript is organized as follows. First, the isothermal polymerization of IBoMA under nitrogen is analyzed. The polymerization rate was fitted with a simple model using values reported in the literature for the initiation and propagation rates. In particular, the propagation rate coefficient ($k_p$) for this monomer was recently obtained using pulsed-laser polymerization combined with analysis of the ensuing polymer molecular weight distribution.[5] The variation of the termination rate coefficient with conversion was fitted using a free-volume model,[6-8] together with a decrease in the efficiency of the initiator decomposition at medium and high conversions.[9] The model provided a conceptual explanation of the shape of the maximum observed in the polymerization rate (gel or Tromsdorff effect).[3,10-14] In a subsequent section, results obtained for the isothermal polymerization of IBoMA in air are discussed. An original finding was the observation of two well-defined maxima, a fact that was particularly evidenced when using small masses of samples (high surface areas per unit volume). To the best of our knowledge, the presence of two maxima in an isothermal rate free-radical polymerization rate, has not been previously reported in the literature. A qualitative explanation of this effect, associated to the specific characteristics of the termination step with oxygen, is provided.

**Experimental Section.**

**Materials.** Isobornyl methacrylate (IBoMA, Aldrich) was used as received (Figure 1). It contained 150 ppm p-methoxy phenol (MEHQ, methylether hydroquinone), as inhibitor. Benzoyl peroxide (BPO, Akzo-Nobel) was used as initiator.

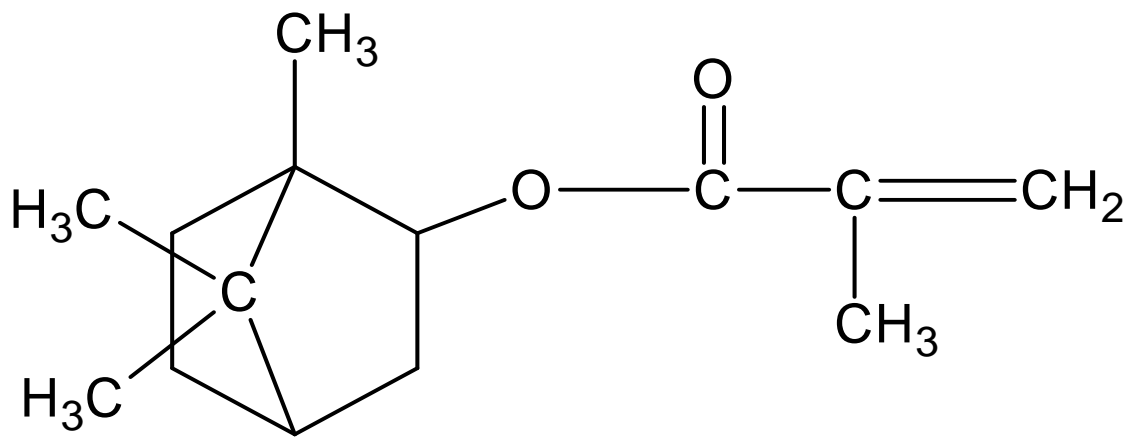

**Figure 1.** Chemical structure of IBoMA.

**Kinetics.** The homopolymerization of IBoMA in the presence of 2 or 3 wt % BPO, was followed at 80 ºC using differential scanning calorimetry (DSC, Pyris 1, Perkin-Elmer or DSC-50 Shimadzu), and Fourier-Transformed Infrared Spectroscopy (FTIR, Genesis II, Mattson).

DSC pans containing the monomer and initiator were previously sealed in air or in a chamber purged with nitrogen. Both isothermal and dynamic runs were performed circulating nitrogen in the DSC cell (outside the sealed pans). Dynamic runs carried out at 5 ºC/min, were employed to determine the overall reaction heat. Isothermal runs enabled to obtain the polymerization rate (proportional to the height of the signal) as a function of time or conversion (proportional to the partial area under the curve).[15] Subsequent dynamic runs at 5 ºC/min were used to determine residual reaction heats.

The kinetic study with FTIR was performed using 3-mm glass tubes to keep temperature constant in the course of polymerization. The very low surface area per unit volume enabled to neglect the influence of oxygen on the polymerization rate. Tubes were filled with the IBoMA-BPO solutions and placed in a thermostat at 80 ºC; they were retired from the thermostat at pre-specified times and quenched in an ice-water mixture to stop the reaction. Partially converted samples were dissolved in methylene chloride (3.5 g in 100 g of solvent) and FTIR spectra were recorded. The monomer conversion was followed by measuring the absorbance of the C=C stretching vibration at 1640 cm$^{-1}$. In order to take into account small variations in the concentration in methylene chloride, the band at 1455 cm$^{-1}$ assigned to a combination of asymmetrical C-CH$_3$ vibrations and to C-H bending in CH$_2$ groups, was used as a reference. By calling, $h = A_{1640}/A_{1455}$, the monomer conversion was defined as:

$$x = 1 - [(h(t) - h_{PIBMA})/(h(0) - h_{PIBMA})], \qquad (1)$$

where $h_{PIBMA}$ takes into account the residual (very small) absorbance at 1640 cm$^{-1}$ present in the polymer.

**Glass Transition Temperature vs. Conversion.** In order to obtain the vitrification curve, solutions containing different concentrations of poly(isobornyl methacrylate) (PIBoMA) in the monomer were prepared. For this purpose, PIBoMA was synthesized by heating about 2 g of an IBoMA-BPO solution containing 2 wt % BPO, placed in a closed glass tube, at 80 ºC during 1 h. The polymerization was completed during 30 min at 140 ºC (no residual reaction heat was observed by differential scanning calorimetry).

Glass transition temperatures ($T_g$) of solutions of the polymer (PIBoMA) in the monomer (IBoMA), were determined from the change in the specific heat measured in dynamic DSC runs at 5 ºC/min (Pyris 1, Perkin-Elmer). Both the onset and end values of the transition were recorded.

**Molar Mass Distributions.** Molar mass distributions of PIBoMA, relative to polystyrene standards, were determined by size exclusion chromatography (SEC, Knauer K-501, RI detector Knauer K 2301, Columns: Phenomenex Phenogel M2 and Waters Ultrasyragel 10$^4$ Å, THF as carrier at 1 ml/min; PIBoMA concentration in THF = 2 mg/ml).

**Results and Discussion.**

**Reaction Heat.** It was obtained from dynamic DSC runs using samples containing 2 wt % BPO. The resulting average value was 54.2 kJ/mol ± 2.7 kJ/mol, in agreement with values reported in the literature for methacrylate polymerizations.[16] Isothermal runs carried out at 80 ºC led to a reaction heat of 43.7 kJ/mol ± 2 kJ/mol, close to 80 % of the total reaction heat. The residual reaction heat measured in a subsequent dynamic run, was comprised in the range of 9 – 12 kJ/mol, giving a total reaction heat consistent with the value obtained in dynamic runs.

The total reaction heat was the same for runs carried out in nitrogen or air. The presence of oxygen affected the polymerization rate (as will be analyzed in the corresponding section), but the reaction continued to complete conversion.

**Vitrification in the Course of an Isothermal Polymerization.** Particular conversions were simulated employing solutions of the polymer (PIBoMA) in the monomer (IBoMA). Figure 2 shows experimental curves of the beginning (onset value) and end of the glass transition of these solutions. For a particular conversion, read in the ordinates, the glass transition range was comprised between the onset and end curves. In the course of a polymerization at 80 ºC, vitrification starts when the system reaches the end curve at a conversion close to 0.76, and is completed at a conversion close to 0.84 (onset curve). Within the sensitivity of the DSC, the polymerization rate became negligible at a conversion close to 0.80, when the system evolved through the vitrification range.

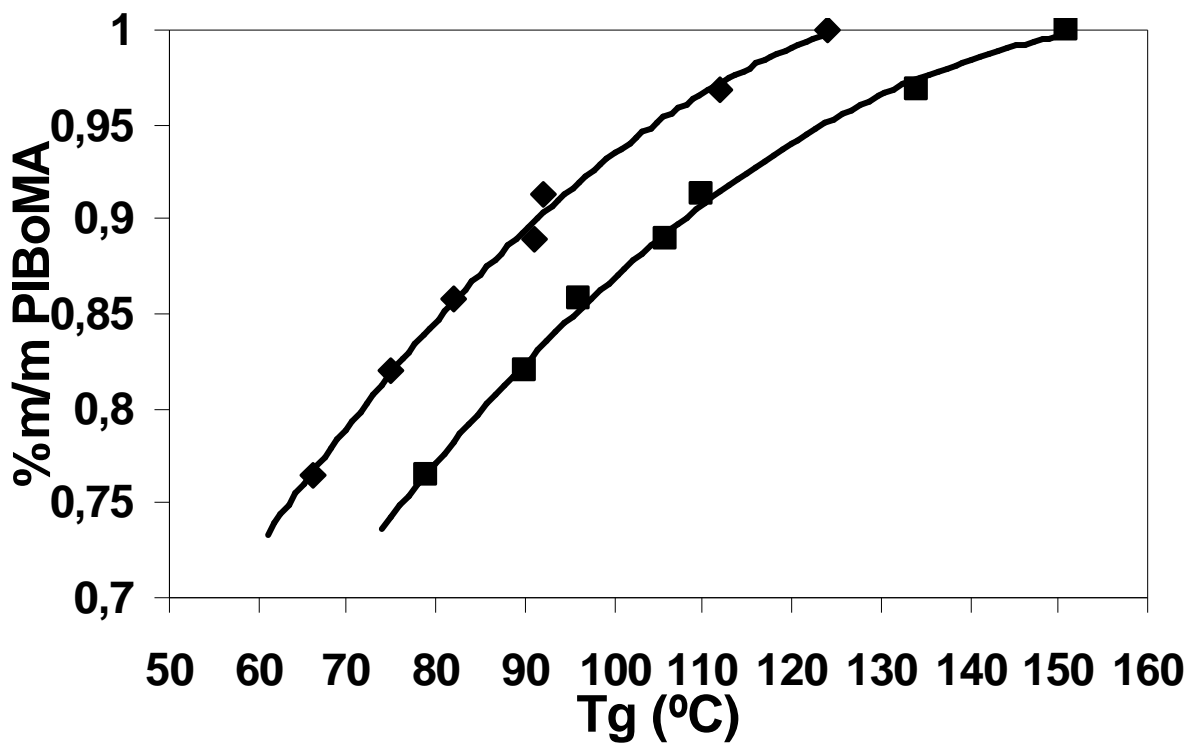

**Figure 2.** Vitrification range of solutions of the polymer (PIBoMA) in the monomer (IBoMA), represented in a conversion (wt % PIBoMA) vs. temperature transformation diagram.

**Polymerization Rate in a Nitrogen Atmosphere.** Figure 3 represents the polymerization rate obtained by DSC at 80 °C, for a sample containing 2 wt % BPO (as a function of time in Figure 3a and as a function of conversion in Figure 3b).

Several facts arise from the inspection of this Figure. The first observation is the presence of an induction period, close to 1 min, where no reaction was observed. Phenol-based inhibitors such as the one present in the commercial sample of IBoMA, are only effective in the presence of oxygen, reacting with $RO_2·$ radicals to form stable products.[17] Therefore, the presence of an induction period must be related to the presence of residual oxygen dissolved in the sample, that was eliminated by reaction with radicals generated by decomposition of the initiator. This means that during the sealing of the pan in a nitrogen atmosphere, oxygen was not completely desorbed from the monomer. This fact was, however, beneficial because it enabled to stabilize temperature and obtain a good baseline before the beginning of polymerization.

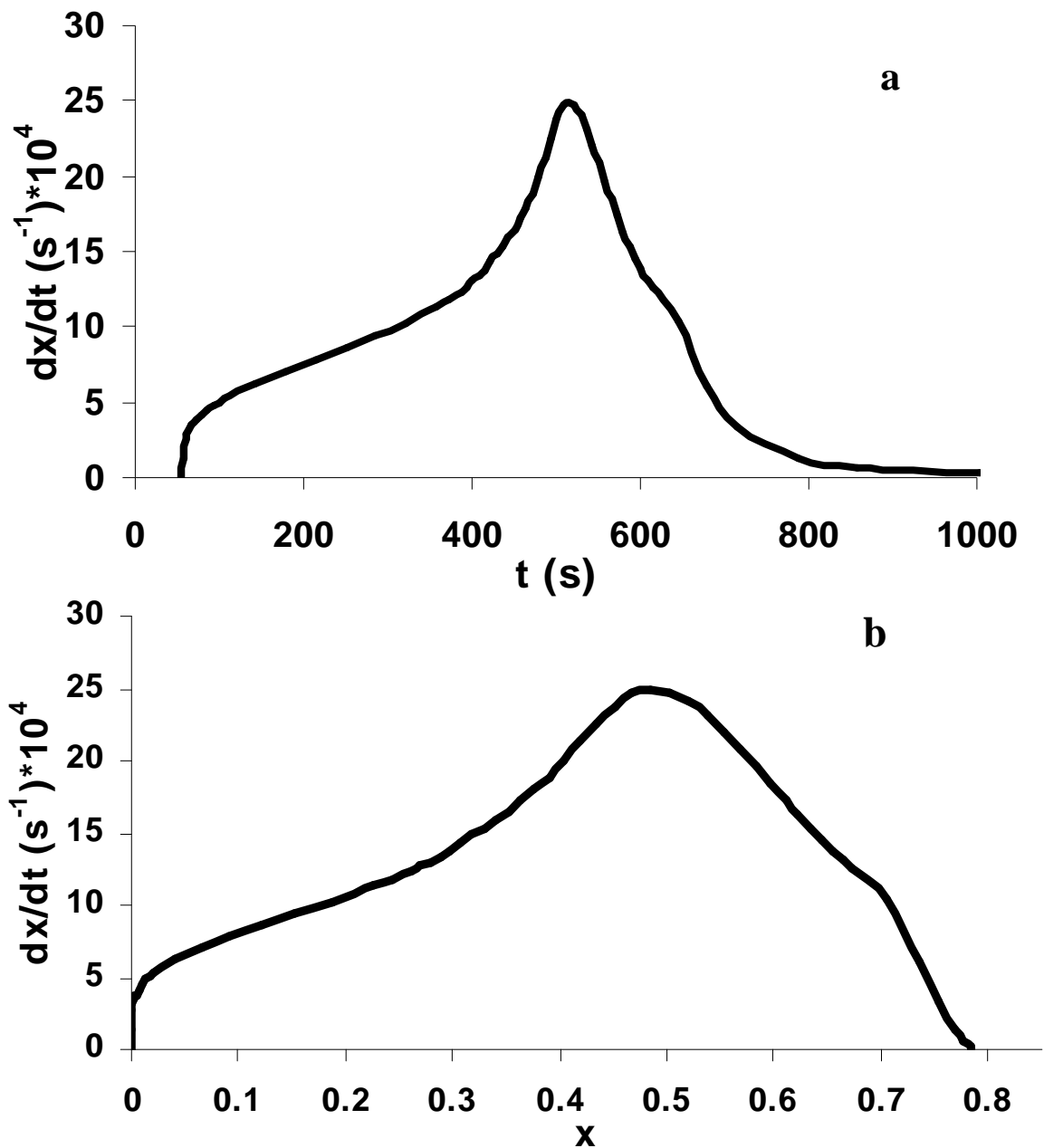

**Figure 3.** Polymerization rate at 80 ºC, in nitrogen, for a sample containing 2 wt % BPO; (a) as a function of time, (b) as a function of conversion.

A second observation is the absence of a plateau in the polymerization rate at low conversions; in fact, the rate increased continuously up to the maximum value (gel effect), attained at a conversion close to 50%. Therefore, there was no particular onset conversion for the gel effect (although at conversions close to 30 % the increase in polymerization rate became more pronounced). This behavior is characteristic of systems where the termination rate is controlled by translational diffusion from the beginning of polymerization.[18-20] The final observation is the fact that the decrease in the polymerization rate became more pronounced at conversions close to 0.7 where the system approached the vitrification region. In this range of conversions a significant decrease of the propagation rate constant may be expected. The polymerization rate became practically negligible at a conversion close to 0.80 (in the middle of the vitrification region).

In order to assess the reliability of the calorimetric information, conversion vs. time curves at 80 ºC for the same formulations, were also obtained by FTIR. Figure 4 shows the comparison of kinetic data obtained from both techniques, represented as a function of time counted after the induction period ($t_i$ ~ 1 min in DSC runs and ~ 3 min in FTIR experiments; the difference is the result of variations in the initial oxygen concentration dissolved in the monomer). A very good matching of both curves was obtained; however, DSC provides much more accurate information of the polymerization rate as a function of conversion.

A simple kinetic model was used to fit the experimental curve of polymerization rate as a function of conversion:

$$- d[I]/dt = k_d[I] \qquad (2)$$

$$d[R^\cdot]/dt = 2fk_d[I] - [2k_t/(1+\varepsilon x)] [R^\cdot]^2 \qquad (3)$$

$$dx/dt = [k_p/(1+\varepsilon x)] [R^\cdot] (1-x) \qquad (4)$$

where $k_d$, $k_p$ and $k_t$, are, respectively, the rate coefficients for initiator decomposition, propagation and termination; $f$ is the efficiency of the initiation process and $\varepsilon$ is the volume expansion factor given by $(\rho_m - \rho_p)/\rho_p$, where $\rho_m$ and $\rho_p$ are the monomer and polymer density, respectively.

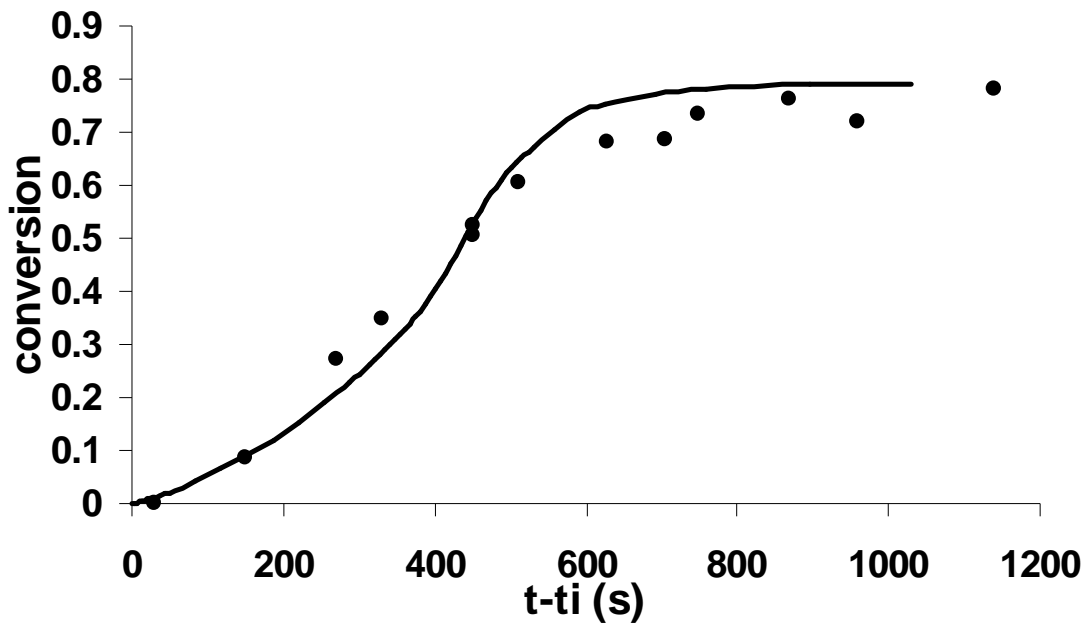

**Figure 4.** Conversion vs. time curves at 80 °C, for samples containing 2 wt % BPO, obtained using DSC (with nitrogen in the headspace; continuous trace), and FTIR (points). Times are counted from the beginning of polymerization (after the induction period).

The volume expansion factor was determined from the experimental values of the densities of monomer and polymer, leading to: $\varepsilon = -0.0755$. Value of $k_d$ for BPO at 80 °C, are reported in the literature for different solvents.[21] They range from $2.5 \times 10^{-5}$ to $5 \times 10^{-5}$ in most solvents. For the present fitting we took $k_d = 4.17 \times 10^{-5}$ s$^{-1}$, giving a half-life of 277 min at 80 °C (reported half-lives at 80 °C were 275 min in the presence of acrylonitrile or styrene and 300 min in the presence of methylmethacrylate).[22] The initial concentration of initiator was $[I]_0 = 0.077$ mol L$^{-1}$ (corresponding to a 2 wt % BPO). Due to the large half-life time of BPO at 80 °C, compared to the polymerization period, the concentration of initiator remained almost constant. The propagation rate constant for this polymerization was recently reported by Hutchinson et al.,[5] combining pulsed-laser polymerization with analysis of the ensuing polymer molecular weight distribution. The value at 80 °C was 2 x 10$^3$ L mol$^{-1}$s$^{-1}$. This value was considered constant up to a conversion $x = 0.7$, where inspection of Figure 3 indicated the beginning of a sharp decrease in the polymerization rate.

Both $k_t$ and $f$ have to be considered a function of conversion. However, the usual finding is that $f$ remains almost constant up to intermediate conversions and only then exhibits a sharp drop.[9,23] Therefore, experimental values of polymerization rate were fitted up to intermediate conversions using a constant value of $f$ and an appropriate function of $k_t(x)$; then, this function was extrapolated (considering the possible appearance of a different termination mechanism), and the value of $f$ was continuously adjusted to fit experimental values of polymerization rate up to a conversion, $x = 0.7$. The initial value of the initiator efficiency was taken as $f = 0.8$ (average of reported values in the range of 0.6 to 1).[22]

The model used to fit the translational diffusion arises from the Fujita-Doolittle theory based on the free volume concept:[6-8]

$$D = D_0 \exp(-C/V_f) \qquad (5)$$

where $C$ is a constant and $V_f$ is the free volume, given by:

$$V_f = \phi_m V_{f,m} + (1-\phi_m)V_{f,p} \qquad (6)$$

The free volume of monomer ($V_{f,m}$) and polymer ($V_{f,p}$) depend only on temperature and may be taken as constants at the isothermal polymerization temperature. The volume fraction of monomer, $\phi_m$, is related to conversion by:

$$\phi_m = (1-x)/(1+\varepsilon x) \qquad (7)$$

Writing eq 5 in the limit of pure polymer leads to:

$$D_p = D_0 \exp(-C/V_{f,p}) \tag{8}$$

When translational diffusion controls the termination rate, the following relationship may be stated:

$$k_t/k_{t,p} = D/D_p \tag{9}$$

Replacing eqs 5-8 in eq 9 and rearranging, leads to

$$\log k_t = \log k_{t,p} + (1-x)/[A(1+\varepsilon x) + B(1-x)] \tag{10}$$

Eq 10 provided an excellent fitting of the polymerization rate in the conversion range 0.04 to about 0.4, using $f = 0.8$ (Figure 5). Values of the adjustable constants were $\log k_{t,p} = 2.219$, $A = 0.1203$ and $B = 0.07046$, for $k_t$ in L mol$^{-1}$s$^{-1}$.

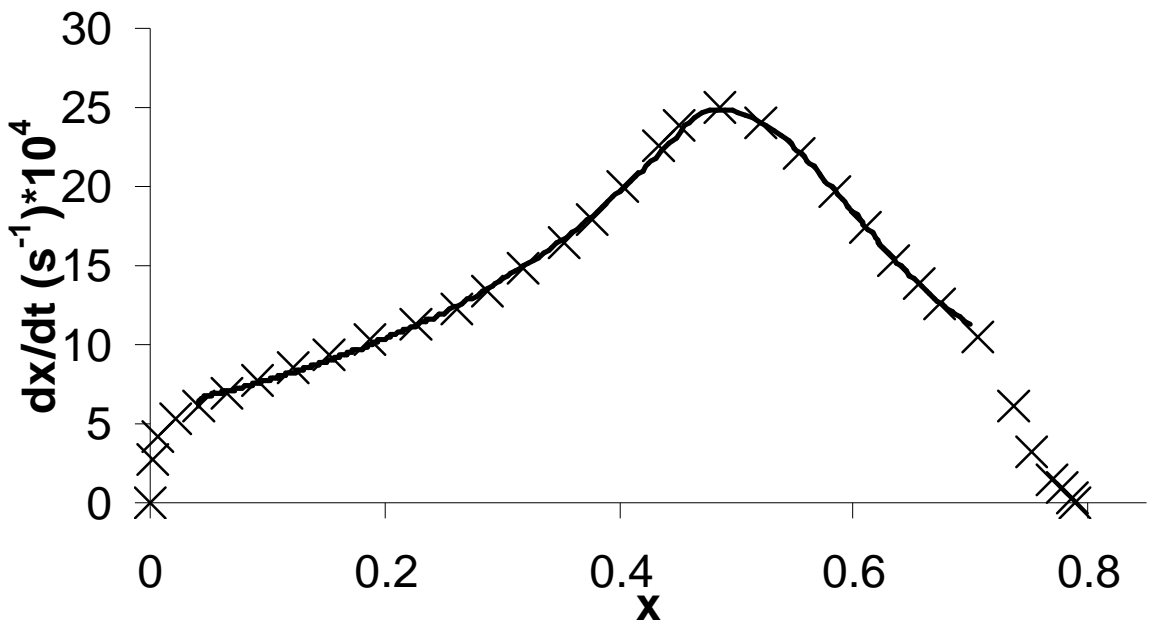

**Figure 5**. Fitting of the experimental polymerization rate (80 °C, under nitrogen, 2 wt % BPO), with the proposed kinetic model, in the 0.04 – 0.7 conversion range (crosses represent experimental values and the full line represents the model prediction).

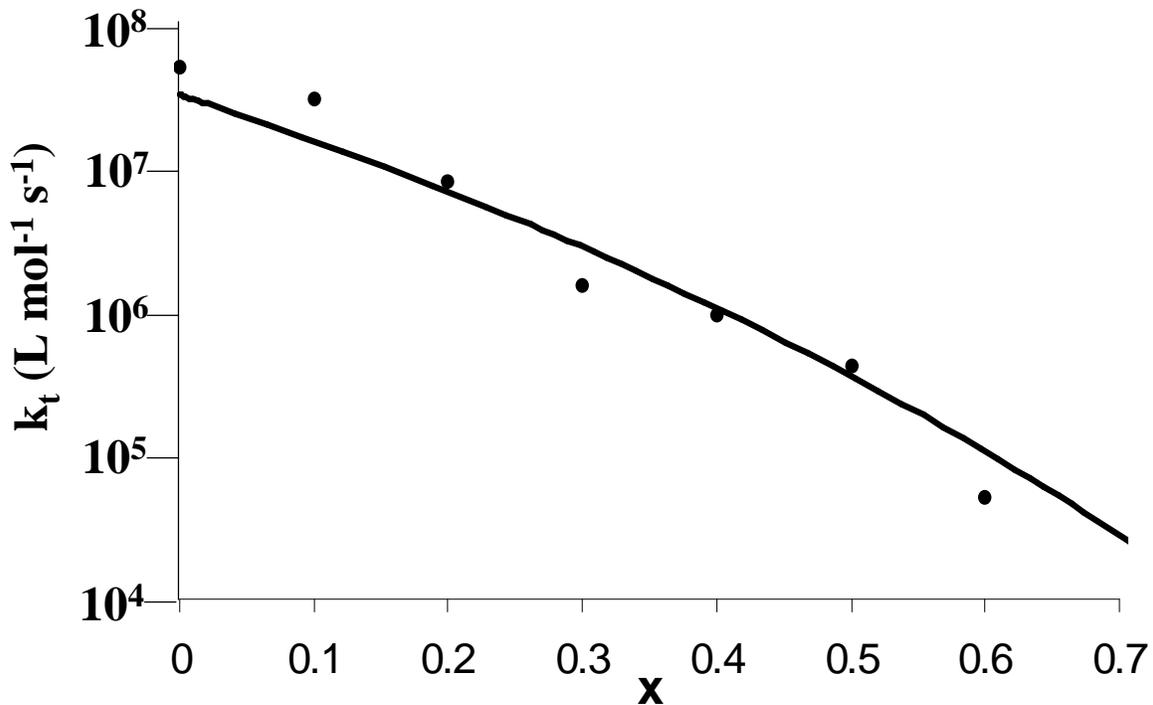

**Figure 6.** Variation of the termination rate coefficient controlled by translational diffusion, with conversion (full curve = function used to fit the present experimental data; points are values reported in the literature for the polymerization of methyl methacrylate at 22.5 °C).[24]

Figure 6 shows $k_t$ as a function of conversion, as predicted from eq 10. The range of $k_t$ values agrees with those reported for the free-radical polymerization of other monomers.[18-20,24] In particular, experimental points reported for the polymerization of methyl methacrylate at 22.5 °C,[24] are plotted to assess the reasonability of the resulting $k_t$ vs. $x$ functionality. However, at high conversions the main termination mechanism may shift from translational diffusion to "reaction-diffusion".[18-20] In this mechanism free-radical sites come into contact through the propagation reaction. The termination rate coefficient by this mechanism ($k_{RD}$) is proportional to the propagation rate coefficient:

$$k_{RD} = C_{RD} k_p [M] \qquad (11)$$

Reported values of $C_{RD}$ were comprised in the range of 2-3 L mol$^{-1}$ for several dimethacrylates,[25] 3-7 L mol$^{-1}$ for a series of acrylate monomers,[26] 9-10 for methyl methacrylate,[27] compared with a theoretical estimation of 5.7, obtained using the Smoluchowski model.[25]

For a conversion $x = 0.7$ ([M] = 1.33 mol L$^{-1}$), the order-of-magnitude of $k_{RD}$ may be estimated using $C_{RD}$ values in the reported range. This leads to $k_{RD} \sim 10^4$ L mol$^{-1}$s$^{-1}$. For the same conversion, $k_t$ extrapolated from eq 10 is 2.5 x 10$^4$ L mol$^{-1}$s$^{-1}$. This means that translational diffusion may be considered as the mechanism controlling termination in the 0 – 0.7 conversion range (although some influence of reaction-diffusion might be present at the end of this range). Therefore, eq 10 was also used to fit the termination rate coefficient in the 0.4 – 0.7 conversion range.

The continuous increase in polymerization rate at low conversions is clearly the result of the decrease in the termination rate constant with conversion, explained in the frame of a free-volume model. But why does a maximum appear? Frequent explanations are related to the following factors: monomer depletion, change in the termination mechanism from translational diffusion to reaction-diffusion and decrease in the initiator efficiency. In the present case the shape of the polymerization rate curve close to the maximum could only be fitted by assuming the sharp decrease in the initiator efficiency shown in Figure 7. A similar behavior was experimentally reported for the bulk polymerization of methyl methacrylate using different initiators.[9,23] As $f$ for different initiators declines rapidly at significantly different conversions,[23] it would be interesting to study the same polymerization using different initiators, to monitor possible changes in the location of the maximum in the rate curves.

Using eq 10 together with $f(x)$ plotted in Figure 7, an excellent fitting of the polymerization rate was obtained for the 0.04-0.7 conversion range, as shown in Figure 5.

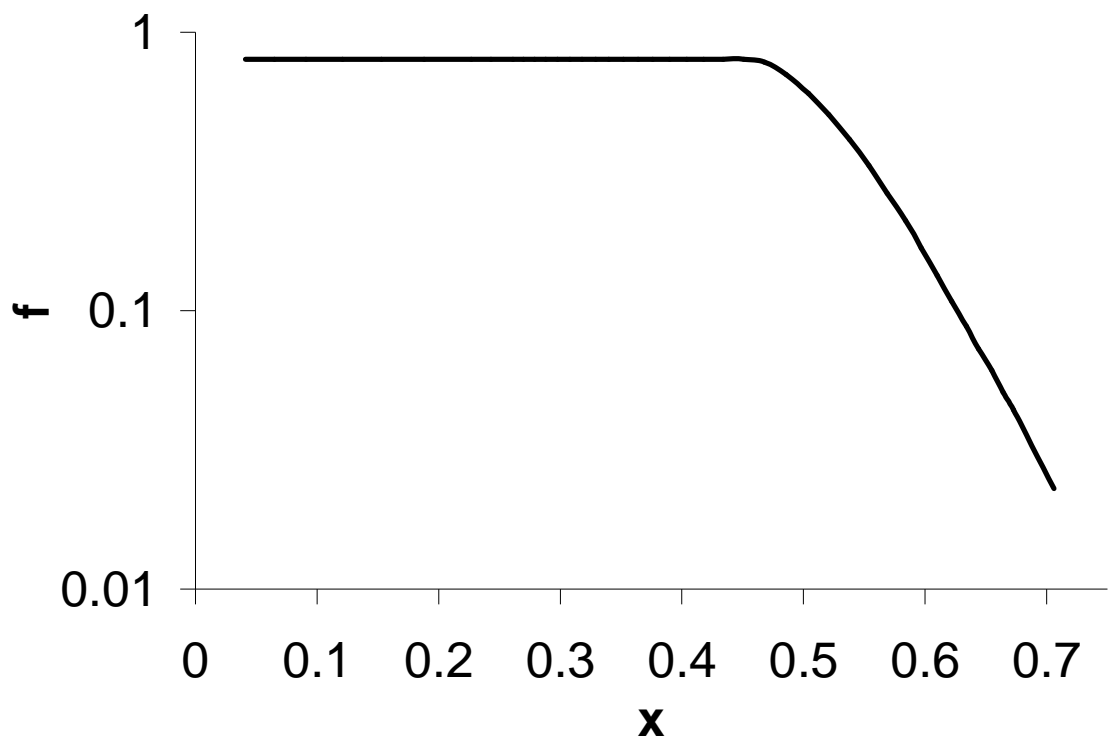

**Figure 7.** Variation of the initiator efficiency with conversion proposed to fit the polymerization kinetics.

**Polymerization rate in air.** A series of polymerization reactions was performed at 80 ºC in DSC pans sealed in air, with different masses of samples containing 2 wt % BPO. This enabled to vary the surface-to-volume ratio of samples exposed to air. Figure 8 shows the set of experimental results.

The first clear effect arising from Figure 8 is the significant decrease in the average polymerization rate when increasing the surface-to-volume ratio of the sample. The retardation caused by oxygen must be effective in a small region close to the gas-liquid interface, as is frequently stated in the literature.[1,19] A second observation is that the final conversion of every sample was close to 0.80, meaning that reaction was arrested by vitrification at the same conversion than the one obtained when polymerization was performed in nitrogen. But, the original finding was the presence of two maxima in the polymerization rate, an effect which was much more evident when decreasing the sample size. This implies that the isothermal polymerization rate experienced a first acceleration at low conversions and a second acceleration at high conversions, followed by the final drop in the vitrification region. Figure 9 shows the same kinetic curves plotted as a function of conversion.

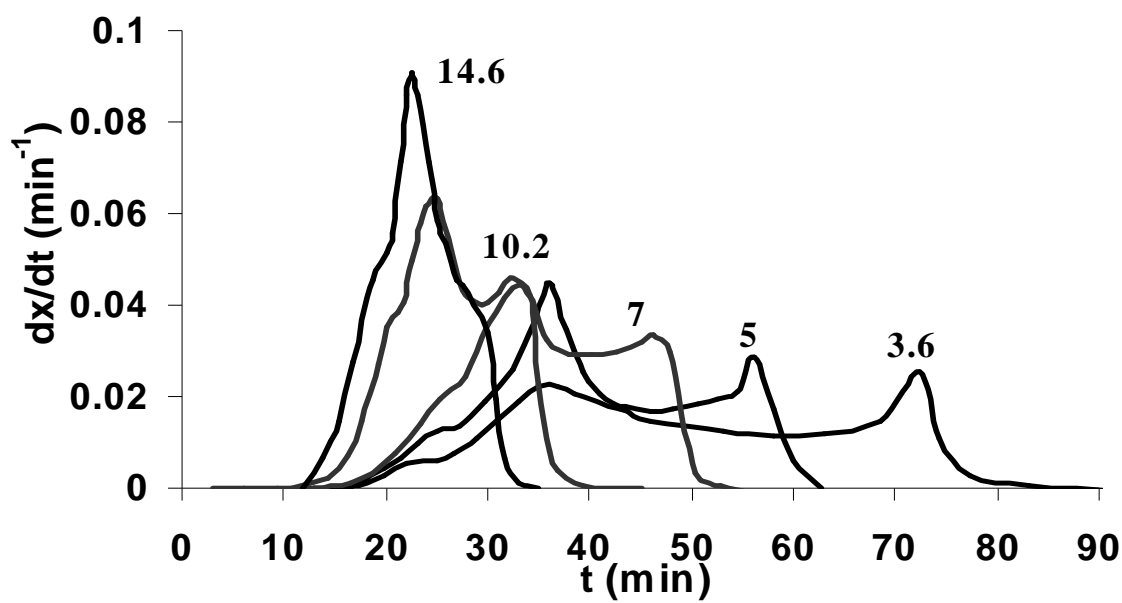

**Figure 8.** Isothermal DSC runs at 80 ºC for the polymerization of IBoMA with 2 wt % BPO, in pans sealed in air (the sample mass in mg is indicated over each curve; DSC Pyris 1, Perkin-Elmer).

In order to verify that the presence of two maxima was not produced by an (unknown) experimental artifact, we performed different experimental tests. First, we check that the phenomenon could be replicated as was indeed the case. Then, we check the constancy of temperature in the course of polymerization (monitoring the corresponding DSC signal), and found it to be constant as expected (small sample sizes combined with low polymerization rates). Next, we change the DSC device (DSC-50 Shimadzu instead of Pyris-1 Perkin-Elmer), and obtained a similar trend. Finally, the amount of BPO was increased from 2 wt % to 3 wt % and the same effect was found (Figure 10). The presence of two maxima in the isothermal reaction rate seems to be a characteristic of samples exhibiting a large surface-to-volume ratio undergoing a free-radical polymerization in air.

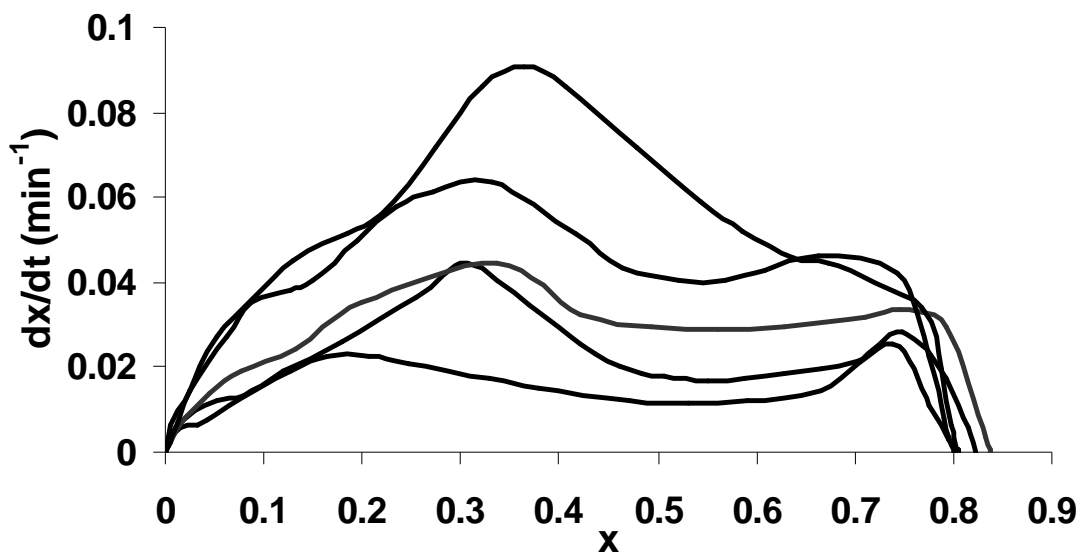

**Figure 9.** Kinetic curves of Fig. 8 plotted as a function of conversion.

In order to provide a qualitative explanation of the presence of two maxima, a run obtained in the DSC-50 Shimadzu, using 4 mg of sample, was selected. The initial thickness of the monomer film in the DSC pan was close to 220 μm. Figure 11 shows the polymerization rate as a function of time (Figure 11a), and conversion (Figure 11b).

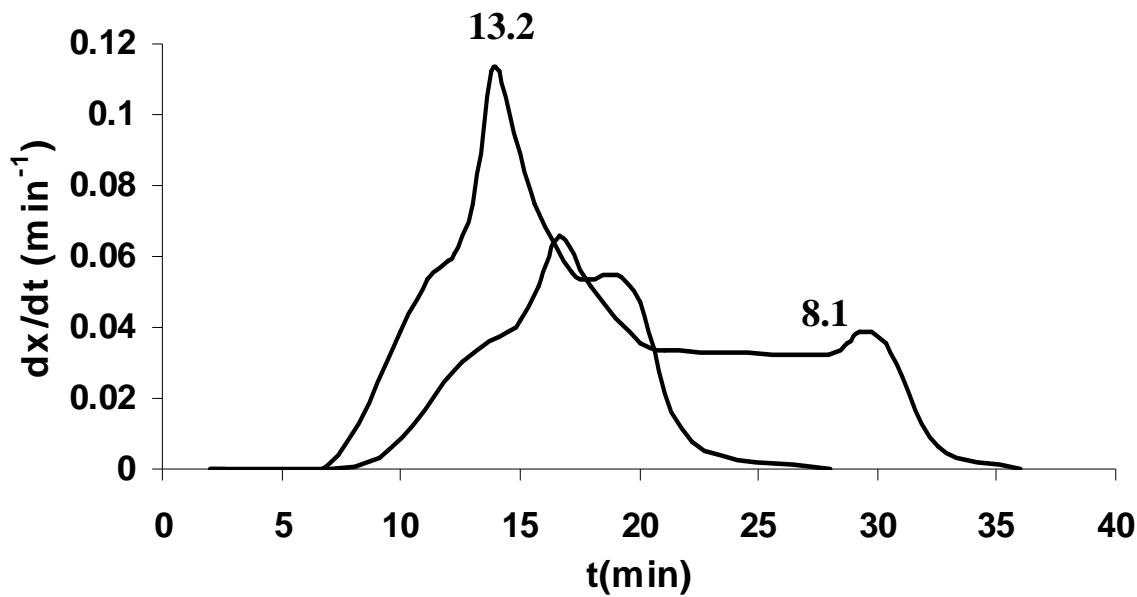

**Figure 10.** Isothermal DSC runs at 80 ºC for the polymerization of IBoMA with 3 wt % BPO, in pans sealed in air (the sample mass in mg is indicated over each curve; DSC Pyris 1, Perkin-Elmer).

A description of the polymerization kinetics retarded by oxygen would need the statement of differential mass balances for every species, with time and a coordinate indicating the distance from the gas-liquid interface, as independent variables. Diffusion coefficients of oxygen, monomer, polymer, primary radicals and macromolecular radicals should be taken as a function of conversion in terms of a free-volume model. The oxygen solubility in the reaction medium (equilibrium condition at the interface), should be also expressed as a function of conversion but its partial pressure in air should remain almost constant due to the small fraction that is consumed by reaction with free radicals. Our intention is just to provide a qualitative explanation of the double maximum in reaction rate, taken the previous concepts into account.

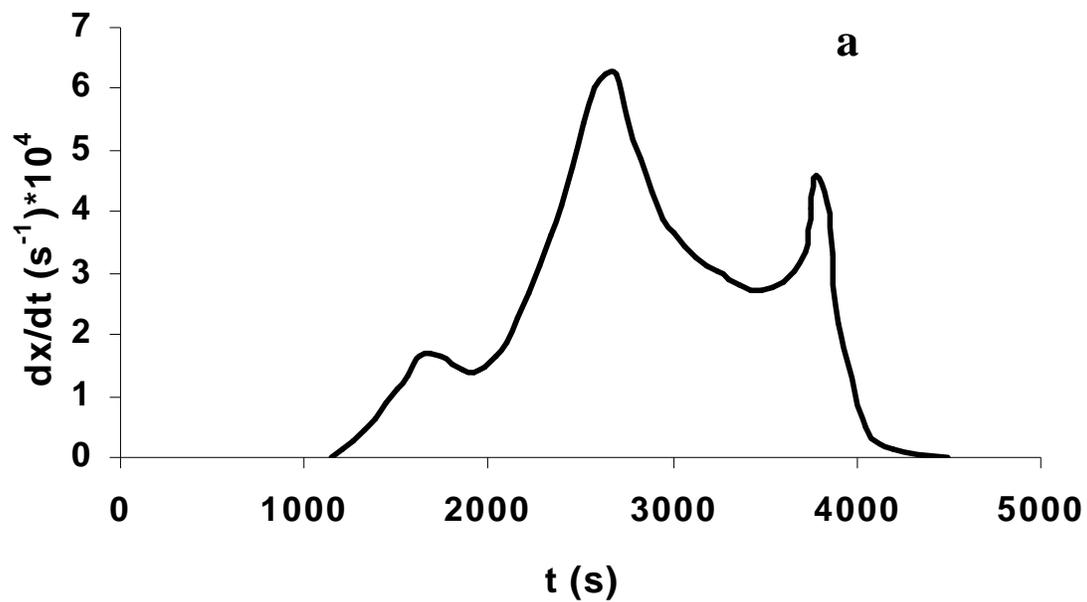

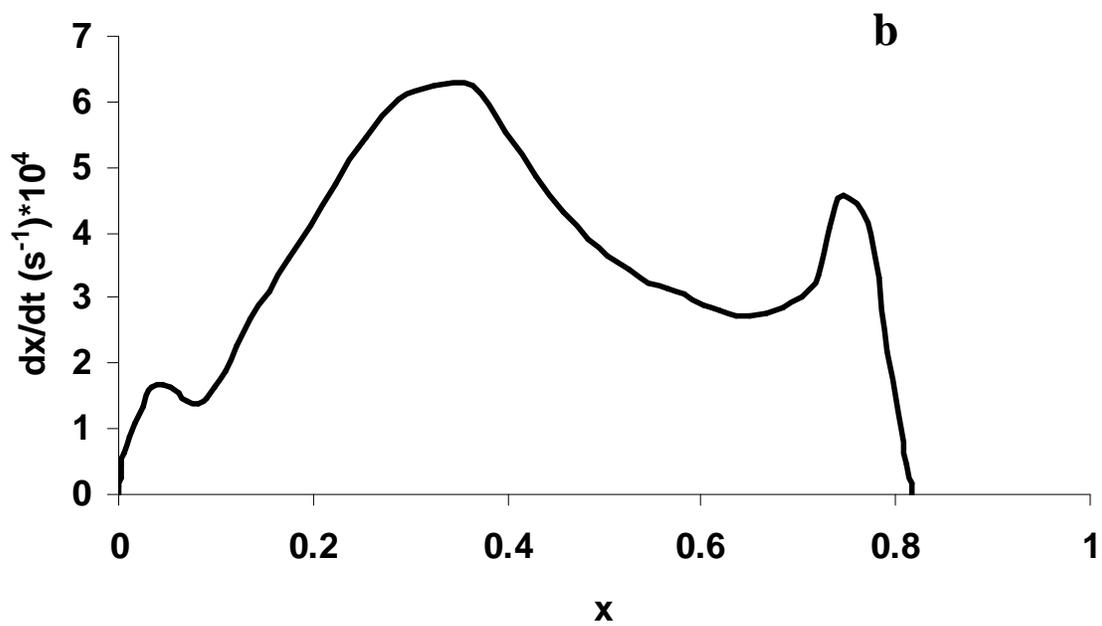

**Figure 11.** Polymerization rate at 80 ºC, in air, for a 4-mg sample containing 2 wt % BPO; (a) as a function of time, (b) as a function of conversion.

First, we will provide an experimental proof of the increase in the termination rate caused by the presence of an oxygen-rich layer close to the interface. This increase is assumed to be produced by diffusion of living radicals to this layer, followed by their fast reaction with oxygen. Due to this new termination mechanism, the number-average molar mass of the polymer obtained in air should be lower than the one synthesized in nitrogen. Size exclusion chromatograms of both polymers, relative to polystyrene standards, are shown in Figure 12. Average molar masses for the polymer synthesized in air were: $M_n = 3.70 \times 10^5$ Da, $M_w = 8.09 \times 10^5$ Da ; corresponding values for the polymer obtained in a nitrogen atmosphere were: $M_n = 5.32 \times 10^5$ Da, $M_w = 1.19 \times 10^6$ Da , in agreement with the expected trend.

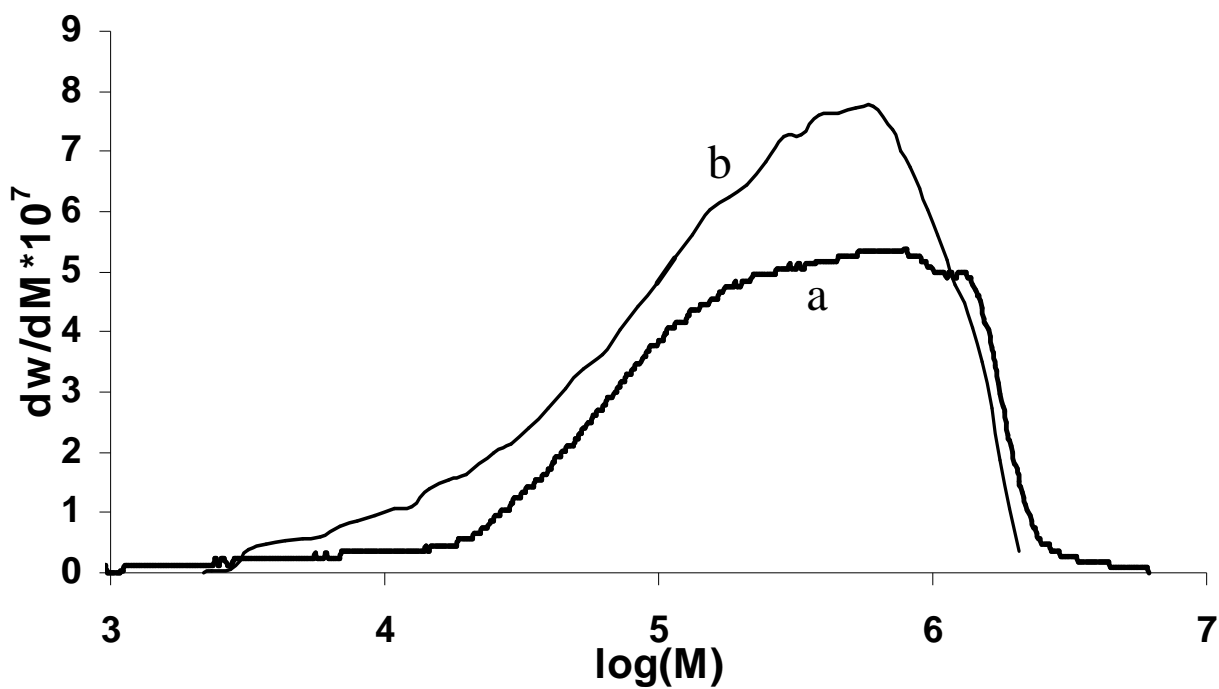

**Figure 12.** Size exclusion chromatograms relative to polystyrene standards, of the polymers synthesized in nitrogen (a), and in air (b).

Figure 11 shows that the polymerization started after an induction period close to 19 min, implying that, at this time, the oxygen concentration at positions distant from the surface was sufficiently depleted. This process of scavenging residual oxygen in the sample continued up to a conversion $x = 0.04$. Then, a short conversion period (up to $x = 0.08$), with an almost constant value of the polymerization rate was observed. This pseudo-steady state period results from the matching of the overall initiation and termination rates, leading to a constant concentration of free radicals. At this early stage of the polymerization, the termination of macromolecular radicals by reaction with oxygen must be very effective due to the relatively high diffusion rates to the interface. Possibly, it is the smallest macromolecular radicals, diffusing more rapidly, that are most frequently terminated by reaction with oxygen during these early stages of the polymerization. In order to test this hypothesis, the molar mass distribution of the polymer obtained at a conversion $x = 0.1$ was compared with the corresponding distribution obtained at the end of polymerization ($x = 0.8$). Size exclusion chromatograms, shown in Figure 13, give evidence of the lower average molar masses of the polymer accumulated at $x = 0.1$ compared to the polymer obtained at the end of the polymerization (as only 0.4 mg of polymer resulted from a single run arrested at $x = 0.1$, it was necessary to repeat the reaction 8 times to obtain enough polymer for a correct characterization).

It is interesting to pay attention to the fact that in the 0.04 – 0.08 conversion range there is in fact a small decrease of the polymerization rate with conversion. This effect may be ascribed to a small increase in the termination rate of free radicals with oxygen, explained by the decreasing size of the polymer coils when increasing their concentration in the monomer (when increasing conversion), and the corresponding increase in the diffusion coefficient.[19]

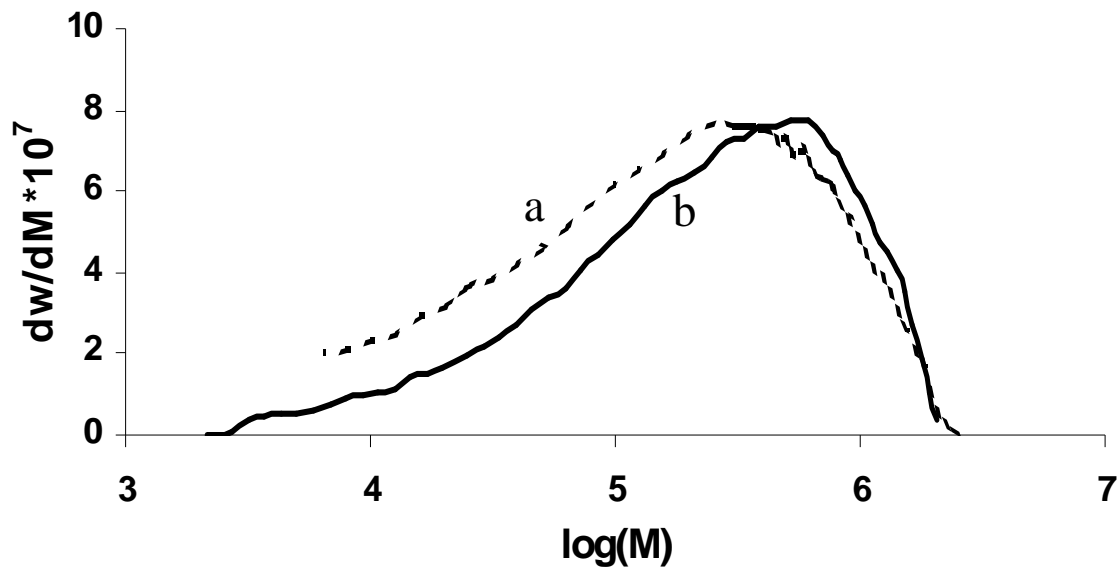

**Figure 13.** Size exclusion chromatograms relative to polystyrene standards, of the polymers obtained at $x = 0.1$ (a) and $x = 0.8$ (b), during the polymerization in air.

The decrease in free volume caused by the increase in conversion, affects both the diffusion of macromolecular radicals to the oxygen-rich boundary and the solubility and diffusion coefficient of molecular oxygen in the reaction medium. The presence of two maxima in the reaction rate may be explained by assuming that constraints to the diffusion of macromolecular radicals start at low conversions (as when the polymerization occurs in the absence of oxygen), and limitations to the solubility/diffusion of oxygen are more severe in the medium-high conversion rage. In fact, it has been reported that radical scavenging by oxygen is retarded by an increase in viscosity and can be severely restricted in the gelled and vitrified states.[19,28] Therefore, there are two different stages where the overall free-radical concentration experiences an increase, leading to the appearance of two maxima in the polymerization rate.

**Conclusions**

The presence of two maxima in the rate of a free-radical polymerization retarded by oxygen is reported for the first time. This finding was confirmed replicating the experiences in two different DSC devices and changing the initiator concentration in the initial formulation. The first maximum was ascribed to the decrease in the diffusion rate of macromolecular radicals to the oxygen-rich boundary, an effect that starts at low conversions. The second maximum was related to the decrease in the solubility/diffusion of oxygen in the reaction medium, a phenomenon that is particularly severe at high conversions. This leads to the presence of two maxima in a free-radical polymerization rate retarded by oxygen.

**Acknowledgements.** The financial support of the National Research Council (CONICET, Argentina), the National Agency for the Promotion of Science and Technology (ANPCyT, Argentina), and the University of Mar del Plata, is gratefully acknowledged.